\documentclass[12pt]{article}
\usepackage{epsfig}
\usepackage{cite}\usepackage{amssymb}
\usepackage{a4}
 \usepackage{comment}
\begin{document}
\bibliographystyle{unsrt}

 \def\jour#1#2#3#4{{#1} {\bf#2} (19#3) #4}
\def\appj#1{{#1}}
\def\jou2#1#2#3#4{{#1} {\bf#2} (20#3) #4}
\def\PL{Phys. Lett.  {B}}
\def\ZP{Z. Phys.  {C}}
\def\EPJ{Eur. Phys. J. {C}}
\def\NPB{Nucl. Phys. {B}}
\def\NPBP{Nucl. Phys. {B} (Proc. Suppl.)}
\def\PRp{Phys. Reports}
\def\PRD{Phys. Rev. {D}}
\def\PRC{Phys. Rev. {C}}
\def\PRL{Phys. Rev. Lett}
\def\IJ{Int. J. Mod. Phys. {A}}
\def\ML{Mod. Phys. Lett. {A}}
\def\JP{J. Phys. {G}}
\def\AP{Acta Phys. Pol. {B}}
\def\NIM{Nucl. Instr. Meth. {A}}
\def\CP{Comp. Phys. Comm.}
\def\Usp{Physics - Uspekhi}
\def\JL{JETP Lett.}
\def\Phy{Physica {A}}
\def\RPP{Rep. Prog. Phys.}
\def\NPYF{Sov. J. Nucl. Phys.}

\def\nwp{\newpage}
\def\nopar{\noindent}
\def\bi{\bibitem}
\def\vs{\vspace*}
\def\hs{\hspace*}
\def\ct{\cite}
\def\beq{\begin{equation}}
\def\eeq{\end{equation}}
\def\bea{\begin{eqnarray}}
\def\eea{\end{eqnarray}}
\def\la{\label}
\def\bec{\begin{center}}
\def\ece{\end{center}}
\def\bef{\begin{figure}}
\def\efi{\end{figure}}
\def\bei{\begin{itemize}}
\def\eit{\end{itemize}}
\def\itm{\item}
\def\epss{\epsfysize}
\def\epsf{\epsffile}
\def\cptn{\caption}

\def\al{\langle}
\def\ar{\rangle}
\def\leq{\leqslant}
\def\geq{\geqslant}
\def\lssim{\stackrel{<}{_\sim}}
\def\gtsim{\stackrel{>}{_\sim}}

\def\ea{{\sl et al.}}
\def\eg{{\sl e.g.}}
\def\et{{\sl etc.}}
\def\ie{{\sl i.e.}}
\def\va{{\sl via }}
\def\vrs{{\sl versus}}

\def\ep{e$^+$e$^-$}
\def\epr{e$^+$p}
\def\z0{${\rm Z}$}
\def\pp{p_{\perp}}
\def\pT{p_{\rm T}}
\def\ptc{\pT^{\,\mathrm{cut}}}
\def\ptcm{p_{T {\rm min}}^{\,\mathrm{cut}}}
\def\pc{p^{\,\mathrm{cut}}}

\def\fptc{F_q(\ptc)}
\def\fpc{F_q(\pc)}
\def\kptc{K_q(\ptc)}
\def\kpc{K_q(\pc)}

\def\HW{{\sc Herwig}}
\def\JT{{\sc Jetset}}
\def\PT{{\sc Pythia}}
\def\ARI{{\sc Ariadne}}
\def\OP{OPAL}
\def\DE{DELPHI}
\def\col{Collaboration}
\newcommand{\etal}{{\em et al.\/}}
\newcommand{\NP}[3]{Nucl. Phys. {\rm {#1}} ({#2}) {#3}}

\def\errfigcap{The error bars, shown where larger than the marker size, 
represent the
statistical and systematic uncertainties added in quadrature.\
}

\def\becfigcap{The dotted line shows Monte Carlo predictions from \PT\ 
with Bose-Einstein
correlations included (see text).\
}

\def\mcerfigcap{The errors of the Monte Carlo predictions are comparable
to those of the data.\
}


\begin{titlepage}
\flushbottom \bec {\large\bf  EUROPEAN ORGANIZATION FOR NUCLEAR RESEARCH} \ece
 \bigskip
\begin{flushright}
{\large  CERN-PH-EP/2006-003}\\
{\large  3rd March 2006}\\
\end{flushright}
\bigskip
\bigskip
\bigskip
\bigskip
 \vspace*{-.7cm}
\bec
\Huge \bf  QCD coherence and correlations
\\  
 \vspace*{-.2cm}
of particles with restricted momenta
\\
 \vspace*{-.2cm}
 in hadronic Z decays 
\ece
\medskip
\bigskip
\bigskip  
\bec
{\LARGE\bf The OPAL Collaboration}
\ece
\bigskip 
\bigskip  
\bigskip
 %
\begin{abstract}

 \nopar
QCD
coherence effects are
studied
based on measurements of
correlations of particles with either
restricted transverse momenta,
$\pT<\ptc$, where $\pT$ is defined with respect to the thrust axis,
or restricted absolute
momenta, $p\equiv |{\bf p}| < \pc $, using about four million 
hadronic \z0\
decays recorded at LEP with the \OP\ detector.
 The correlations are analyzed in
terms of normalized factorial and cumulant moments. The analysis is
inspired by analytical QCD calculations which, in conjunction with
Local Parton-Hadron Duality (LPHD), predict that, due to colour
coherence, the multiplicity distribution of particles with
restricted transverse momenta 
 should 
 become Poissonian as $\ptc$ 
decreases.
The
expected correlation pattern is indeed observed
 down to $\ptc \approx 1$~GeV
but not at lower transverse
  momenta. 
 Furthermore, 
for $\pc\to0$~GeV a strong rise is 
observed in
the data, in disagreement with theoretical expectation.
     The Monte Carlo models reproduce well the
measurements at large
$\ptc$ and $\pc$ but underestimate their magnitudes at the lowest momenta.
The \ep\ data are also compared to the measurements in
deep-inelastic \epr\ collisions.
Our study indicates difficulties
with the LPHD hypothesis when applied to
many-particle inclusive observables
of soft hadrons.
 \end{abstract}
\vspace{1.cm} 
  \vspace*{.5cm}
\centerline {\Large {\it   Submitted to  Physics Letters B}}

\end{titlepage}

\newpage
 \begin{center}{\Large        The OPAL Collaboration
}\end{center}\bigskip
\begin{center}{
G.\thinspace Abbiendi$^{  2}$,
C.\thinspace Ainsley$^{  5}$,
P.F.\thinspace {\AA}kesson$^{  3,  y}$,
G.\thinspace Alexander$^{ 21}$,
G.\thinspace Anagnostou$^{  1}$,
K.J.\thinspace Anderson$^{  8}$,
S.\thinspace Asai$^{ 22}$,
D.\thinspace Axen$^{ 26}$,
I.\thinspace Bailey$^{ 25}$,
E.\thinspace Barberio$^{  7,   p}$
% ...
}\end{center}\bigskip
\bigskip
$^{  1}$School of Physics and Astronomy, University of Birmingham,
Birmingham B15 2TT, UK
\newline
$^{  2}$Dipartimento di Fisica dell' Universit\`a di Bologna and INFN,
I-40126 Bologna, Italy
\newline
$^{  3}$Physikalisches Institut, Universit\"at Bonn,
D-53115 Bonn, Germany
\newline
% ...
$^{  5}$Cavendish Laboratory, Cambridge CB3 0HE, UK
\newline
% ...
$^{  7}$CERN, European Organisation for Nuclear Research,
CH-1211 Geneva 23, Switzerland
\newline
$^{  8}$Enrico Fermi Institute and Department of Physics,
University of Chicago, Chicago IL 60637, USA
\newline
% ...
$^{ 21}$Department of Physics and Astronomy, Tel Aviv University,
Tel Aviv 69978, Israel
\newline
$^{ 22}$International Centre for Elementary Particle Physics and
Department of Physics, University of Tokyo, Tokyo 113-0033, and
Kobe University, Kobe 657-8501, Japan
\newline
% ...
$^{ 25}$University of Victoria, Department of Physics, P O Box 3055,
Victoria BC V8W 3P6, Canada
\newline
$^{ 26}$University of British Columbia, Department of Physics,
Vancouver BC V6T 1Z1, Canada
\newline
% ...
$^{  p}$ now at The University of Melbourne, Victoria, Australia
\newline
% ...
$^{  y}$ now at CERN
\newline
% ...
\newpage

 %
%
\pagestyle{plain} \setcounter{page}{1}

\section{Introduction\label{introduction}}

 At high energies, the
annihilation process \ep$\to\: hadrons$ proceeds through
the creation of a highly virtual primary quark and anti-quark  which
initiate 
 a cascade of partons through successive parton
emissions. The evolution of such a parton cascade is well understood
in perturbative Quantum Chromodynamics (QCD) for virtualities,
$Q^2$, of the daughter partons larger  than $Q^2_0$. Here $Q_0$ is
a virtuality cut-off below which the strong coupling constant
becomes large and perturbative methods cease to be valid.
In comparisons to experimental data, $Q_0$ was found to be of the
order
of hadronic masses ($Q_0\sim $
few hundred MeV).

A fundamental property of a QCD cascade, which follows from the
non-Abelian structure of QCD, is colour coherence. This
induces
an angular ordering of subsequent emissions in
the branching process~\cite{angoder} which restricts the phase space
for each subsequent
  parton in the cascade. Angular ordering has important
consequences of which we mention only a few
(see~\cite{doksh:91a,revo,revid,QCDExpTh} for a comprehensive review).
Compared to a
cascade without angular ordering, the single-parton inclusive
distribution is suppressed for soft, or low-momentum,  particles (the 
``hump-back''
plateau)~\cite{lphda}, the mean parton multiplicity evolves less
rapidly with increasing jet energy, and
the rapidity distribution
becomes flat and energy independent for partons with very 
small
transverse momenta~\cite{revo}.

It is remarkable that inclusive characteristics of {\em hadrons}
measured in a variety of hard processes indeed show a behaviour
similar to that expected from perturbative parton-level
calculations~\cite{revo,revid,QCDExpTh}. This indicates that perturbative
QCD
effects and colour coherence in particular leave their imprint on
the hadronic final state 
 even
for quantities which are not infra-red
safe, such as particle multiplicity.  The hypothesis of Local
Parton-Hadron Duality (LPHD)~\cite{lphda} embodies these
observations. According to LPHD, parton-level QCD predictions are
applicable to sufficiently inclusive hadronic observables without
the need for a hadronisation phase: hadronic spectra are
proportional to those of partons if the cut-off $Q_0$ is decreased
towards a small value of the order of $\Lambda$, the QCD-scale.

 Within the LPHD picture, perturbative QCD calculations have been carried out
in the Double Leading Logarithmic
Approximation (DLLA)
 or in the
Modified Leading Logarithmic Approximation (MLLA) 
 which includes
 terms of
order
$\sqrt{\alpha_{\rm S}}\:$  in the strong coupling constant
\ct{doksh:91a,revo}. The
DLLA
calculations neglect energy-momentum conservation in gluon splittings
which is partly taken into account in the MLLA. Although analytical
calculations provide much valuable physical insight, they have often
to be considered as qualitative. More quantitative results are
obtained  from parton-shower Monte Carlo models, the physics
implementation of which strongly resembles the analytical
calculations, but which impose energy-momentum conservation and
include the complete parton-splitting functions.

In spite of its success with single-particle inclusive spectra,
earlier studies have shown that  the applicability of LPHD is less
evident for the moments of single-particle densities at HERA
energies~\cite{mult} and angular correlations  at LEP
\cite{angulee} and at HERA~\cite{angul}. It is
therefore of considerable importance to further test multiparticle
aspects of perturbative QCD predictions, which are sensitive to colour
coherence,  in conjunction with LPHD.

Sensitive studies of colour coherence
were suggested in \cite{ptp} within DLLA
calculations and using factorial moment and cumulant techniques. 
 In a QCD cascade the presence of one gluon enhances the
probability for further gluon emissions,
 causing
positive
correlations. The multiplicity distribution of partons in a jet is
therefore generally broader than a Poisson distribution (which
corresponds to uncorrelated production)  and obeys asymptotic
KNO-scaling~\cite{KNO}.
However,
 it was pointed
out
in \cite{ptp}
that, due to colour coherence, gluons produced with
bounded transverse momenta, $p_T<\ptc$,  where $p_T$ is defined with
respect to the
primary
parton
in a jet,
become, for small  $\ptc$, 
independently emitted from the primary parton.
This implies that their multiplicity distribution becomes Poissonian,
analogous to that of soft photons radiated from a charged particle in QED.
In contrast, for gluons with bounded absolute momenta, $p
\equiv |{\bf p}| <\pc$,
for which the angular ordering constraint is less important, the
distribution remains non-Poissonian even for very small $\pc$.

The DLLA analytical predictions were first tested in
deep-inelastic
\epr\
scattering
by the ZEUS experiment at HERA~\cite{ptpz} using a sample of
$\simeq7500$ high-$Q^2$ events.  Because of low
 statistics, factorial
cumulants were not studied. The factorial moments were measured in
the current region of the Breit frame \cite{Br}. 
 From the significant
discrepancies between data and both the DLLA calculations
and
parton-level {\ARI}\ Monte Carlo \ct{ar41} expectations, the authors 
conclude that the
LPHD hypothesis is strongly violated for many-particle observables.
Monte Carlo models, which include hadronisation effects, reproduce
the correlation pattern of the hadronic final state, although
sizeable discrepancies remain 
 for small values of $\ptc$ and $\pc$.

In this paper, we report the first results in \ep\ annihilation
on factorial moments and cumulants for hadrons with restricted
transverse and absolute momenta in a jet. The measurements are based
on a data sample of about four million \z0\ hadronic decays recorded with
the OPAL detector at the
LEP \ep\ collider at CERN.

 \section{Analysis}
\label{analysis}

 The calculations in \cite{ptp} use factorial
moments and
cumulants known 
to provide a sensitive tool
to probe
multiparticle correlations  \ct{revo,revid,rev1,WE_kniga}.

The normalised factorial moment of order $q$ in a region of phase
space of size $\Omega$  is defined as \beq F_q(\Omega)={ \al
n(n-1)\cdots (n-q+1)\ar} \bigg / {\al n \ar ^q}\:, \quad q\geq 1.
\label{fv}
\eeq
 Here $n$  is the number of particles in $\Omega$ and the angle
brackets $\al \cdots \ar $ denote the average over events. For
uncorrelated particle production within $\Omega$ one has $F_q=1$ for
all $q$.
   The factorial moments describe {\it many}-particle distributions
 via the
relation
$\al n(n-1)\cdots
(n-q+1) \ar=\int_{\Omega}\rho_q(p_1,\dots,p_q)\,\prod_{i=1}^q
d\,p_i$
between
the unnormalised factorial moments in a region $\Omega$ and the inclusive
$q$-particle densities, $\rho_q(p_1,\ldots ,p_q)$, of particles
with
momenta $p_i$.

The normalised factorial cumulants, $K_q(\Omega)$,  or cumulants for
short~\cite{Mue71,cumulants,cumulantsAnbd}, are related to the
factorial moments $F_q(\Omega)$ through the following relations:
 \beq
K_2=F_2-1,\quad K_3=F_3 - 3\,F_2 +2, \quad
K_4=F_4-4\,F_3-3\,F_2^2+12\,F_2- 6.\la{kvfv}
\label{fc}
\eeq
 By construction, $K_q$   is a measure of {\it genuine} multiparticle
{correlations}: $K_q$, 
 representing the correlation function averaged over the region $\Omega$, 
vanishes whenever any one of the $q$ particles
is statistically independent of the others.
 For
uncorrelated particle production, or 
 Poissonian emission, within 
$\Omega$ one has $K_q=0$ for
$q>1$.

The normalised factorial moments of the multiplicity distribution of
gluons which are restricted in either transverse momentum
$p_{T}<\ptc$ (cylindrically-cut phase space) or absolute  momentum
$p
<\pc$ (spherically-cut phase space) are
predicted
  to have the following qualitative behaviour
\cite{ptp}:
    \beq
\fptc \simeq 1+\frac{q(q-1)}{6}\,
                  \frac{\ln (\ptc/Q_0)}{\ln (E/Q_0)}
\quad {\rm for} \quad \ptc \to Q_0, \la{fpt} \eeq
 and
    \beq
\fpc   \simeq 
 C_1(q)
>1
 \quad {\rm for} \quad \pc \to 0\:\: {\rm GeV},
 \la{fp}
 \eeq
where $E$ is the energy of the initial parton, and transverse
momentum is defined with respect to its direction.
 Here and below the $C$-functions are 
 $q$-dependent constants. 
 The cumulants (\ref{kvfv}) are predicted to behave as
\beq \kptc \propto
                  \left( \frac{\ln (\ptc/Q_0)}{\ln (E/Q_0)}\right )^{q-1}
\quad {\rm for} \quad \ptc \to Q_0. \la{kpt} \eeq
 Equation~(\ref{kpt}) shows that the lower-order cumulants dominate
at small $\ptc$.  The spherically-cut cumulants are predicted to
 behave 
 similarly to factorial moments, Eq.~(\ref{fp}): 
    \beq
\kpc  \simeq 
 C_2(q)
>0
 \quad {\rm for} \quad \pc \to 0\:\: {\rm GeV}.
 \la{kp}
 \eeq
 Equations~(\ref{fpt})-(\ref{kp}) illustrate the different influence
angular ordering has on the multiplicity moments and correlations.
Cylindrically-cut moments show positive correlations but approach
the Poisson limit 
 as $\ptc$ approaches $Q_0$.
 On the other hand, for soft gluons with limited absolute momenta,
$p<\pc$, the multiplicity distribution remains broader than a
Poisson distribution for any (small) value of $\pc$.

In~\cite{ptp}, the analytical results have been 
 tested 
at parton
level using the shower Monte Carlo program {\ARI}~\cite{ar41}. The
moments $\fptc$ indeed show the expected decrease for small values
of the cut, $\ptc\le$ 4 GeV; however, they do not fully reach the
Poisson value for $\ptc\to Q_0$ but saturate at values somewhat
larger than one. On the other hand, $\fpc$ moments attain a maximum
around $\pc\approx2$~GeV, and then decrease towards finite values
much above unity for $\pc\to0$~GeV showing that in spherically-cut phase
space there is no Poisson regime.

\section{Experimental details\la{data}}

\subsection{The OPAL detector\la{detector}}

The OPAL detector, operated from 1989 to 2000 at LEP,  is described in
detail elsewhere~\cite{bib-opal}.
The results presented here are mainly based on the information from
the tracking system, which consisted of a silicon microvertex
detector, an inner vertex chamber, a jet chamber with 24 sectors
each containing 159 axial anode wires, and outer $z$-chambers to
improve the $z$ coordinate resolution\footnote{OPAL uses the
right-handed coordinate system  defined
 with the positive $z$  along the direction of the e$^-$ beam and
 the positive $x$ axis pointing towards the centre of the LEP ring.
$r$ is the coordinate normal to the beam axis, $\varphi$  the
azimuthal angle with respect to the $x$ axis,
 $\theta$  the polar angle with respect to the $z$-axis.}.
The tracking system was located in a 0.435~T axial magnetic field and
measured $\pp$, the track momentum transverse to the beam axis, with
a precision of $(\sigma_{\pp} / \pp) = \sqrt{ (0.02)^2 +
(0.0015 \, \pp)^2 }$ ($\pp$
 in GeV)
for \mbox{$|\cos \theta| < 0.73$}.

\subsection{Data and Monte Carlo samples\la{events}}

This analysis is based on a data sample of
 approximately
$3.9$$\times$$10^6$ hadronic \z0\ decays collected with the \OP\
detector 
 between
1991 
 and 1995. About 91\% of this sample 
was taken
 close to the peak of
  the \z0; the remaining part has a centre-of-mass 
energy,
$\sqrt{s}$, within $\pm 3$ GeV of the \z0\ peak.

Further selection criteria are based on a hadronic event selection
procedure described in detail in \cite{opal}. For each event
tracks were accepted only if they had at least 20 measured
points in the jet chamber, the first hit  closer than 70 cm to
the beam axis, the measured closest distance to the
 \ep\ collision point  less than 5 cm in the plane perpendicular to
the beam axis and less than 40
cm along the beam axis, $\pp> 0.15$ GeV,
and $|\cos \theta|<0.94$.

The  event was then  required to  have at least five
 tracks, a momentum imbalance, defined as the magnitude of the vector
sum of momenta of all charged particles,  below 0.4$\sqrt{s}$, a
total energy of the tracks (assumed to be pions) greater than
0.2$\sqrt{s}$, and
 $|\cos \theta_{\rm thr}| <0.9$, where $\theta_{\rm thr}$ is the
polar angle of the event thrust axis with respect to the beam direction
calculated using all tracks as well as electromagnetic and hadronic
calorimeter clusters.
 These criteria provide rejection against background from
non-hadronic Z decays, two-photon and beam-wall interactions,
beam-gas scattering, and ensure that the event is well
contained inside the detector.
A total of
about 2.9 million events
remain after the selection has been applied and are used for
further analysis.

The kinematic variables used in the analysis are defined with
respect to the 
 event 
thrust axis.
 To remain consistent with the theoretical calculations and with similar
 measurements in deep-inelastic scattering, factorial moments and 
 cumulants
 are calculated for all charged particles in a single event-hemisphere,
 defined by a plane perpendicular to the event thrust axis. The particles
 are assigned positive rapidity, 
 $y=0.5\ln [(E+p_{\rm L})/(E-p_{\rm L})]$, with $E$ and $p_{\rm L}$ 
the energy
(assuming the pion mass) and longitudinal momentum component of the
particle,
 and 
 a single randomly chosen
                hemisphere for each event
is used in the analysis.

To correct the measured factorial and cumulant moments for the
effects of detector response, initial-state radiation, resolution
and particle decays, we apply the correction procedure  adopted in
our earlier studies~\ct{Oi,Og}.
 Two samples of more than three million multihadronic events each were
used, generated with the \JT\
7.4/\PT\ 6.2 Monte Carlo model \ct{js74}.
 The first sample does not include the effects of initial-state radiation,
and all particles with lifetimes longer than 3$\times$$10^{-10}$~s
were considered to be stable.
    The generator-level factorial moments, $F_q(\Omega)_{\rm gen}$,  are calculated directly
from the charged particle multiplicity distributions of this sample
without any selection criteria.
 The second sample was generated including the effects of finite lifetimes
and initial-state radiation and was passed through a full simulation
of the \OP\ detector \ct{detsim}.
 The corresponding detector-level moments, $F_q(\Omega)_{\rm det}$,   are 
calculated from
this set using the same reconstruction and selection algorithms as
used for the measured data.
 The corrected moments are then determined by multiplying the measured
ones by the correction factors
 $U_q(\Omega)=F_q(\Omega)_{\rm gen}/F_q(\Omega)_{\rm det}$.
 The correction factors vary between about 0.85 and 1.2.

As systematic uncertainties, we include the following contributions:
 \bei
 \itm
 The statistical error on the correction factors  $U_q(\Omega)$:
 due to the finite statistics of the Monte Carlo samples these are
comparable to those of the data.

\itm Track and event selection criteria variations as in \ct{Oi,Og}.
The moments have been computed changing in turn the following
selection criteria: the first measured point was required to be
closer than 40 cm to the beam,
the momentum was
required to be less than 40 GeV,
the track polar angle acceptance was changed to $|\cos \theta|<0.7$.
 The changes in the corrected moments when the analysis is performed
 with these cuts were taken as systematic uncertainties.
These
changes modify the results by no more than a few percent in the
smallest phase space regions and do not affect the  conclusions.

\itm 
Resonance decays: We
have repeated our calculations
 with the \PT\ 6.2 Monte Carlo model where no decays
 of resonances were
allowed.
 The difference 
 between 
these calculations
  and
those based on  the sample
generated  including the  resonance decays 
 are taken as systematic uncertainties and
do not exceed 2\%.

\itm Two--jet 
selection criteria:  
 For comparison with ZEUS data
\ct{ptpz} where the rate of hard jet production is lower than in \ep\
annihilation at LEP, we have calculated the moments for two-jet events
selected via a thrust value cut
as given in  Sect.~\ref{zeus}.
 Therefore,
 for the results presented in Sect.~\ref{zeus} only, we have repeated
calculations
 for
 two-jet events using
the Durham jet finder~\ct{Durham}.
 We apply this algorithm  with the jet resolution
parameter $y_{\rm cut}=0.03$,  shown~\ct{oycut} to result in well
separated jets while still yielding reasonable event statistics. The
 changes in the 
results based on the two selection methods 
  are taken as systematic uncertainties. 
 These results
agree to within 7\%.

\itm \HW\ based correction factors $U_q$: The correction factors
$U_q(\Omega)$ were derived from  samples generated with the \HW\
Monte Carlo~\cite{hw62}. 
 The differences compared to \PT\ 
   are taken as systematic uncertainties and
do not
exceed 10\%.
 \eit

The total errors have been calculated by adding the systematic and
statistical uncertainties in quadrature and are
 therefore correlated bin-to-bin.
It was further verified that our conclusions remain unchanged when
events taken at energies off the \z0\ peak are excluded from  the
analysis.

The data are compared to model predictions calculated using the
following Monte Carlo generators:
 \bei \itm \PT\  version 6.2 \cite{js74} with the parton shower
followed by string hadronisation. \itm \PT\  version 6.2, as above,
but including the effect of Bose-Einstein correlations. These are
simulated using the BE$_{32}$ algorithm~\ct{be32} implemented in
{\tt PYBOEI}. In a previous OPAL study of higher-order
cumulants~\ct{Og} it was shown that this model accounts
simultaneously for the magnitude and bin-size dependence of
cumulants of like-sign as well as of all-charge multiplets in one-
to three-dimensional phase space. Here we use a Gaussian
parametrisation with {\tt PYBOEI} and QCD/fragmentation parameters
from~\ct{opal-wwbec}.
\itm \ARI\ version~4.1 \cite{ar41} with the colour dipole model for
the parton shower followed by fragmentation as in \JT/\PT. \itm \HW\
version 6.3 \cite{hw62} with a parton shower followed by cluster
fragmentation. \eit

 Each Monte Carlo sample consists of more than three million
events. The simulation parameters of the \JT/\PT\ and \HW\ models
have been tuned to \OP\ data in~\ct{otune-jthw}. The parameters of
\ARI\ and recent changes for the \HW\ parameters\footnote{Here we
used {\tt PSPLT(1)}$\, = 0.6$ and {\tt CLMAX}$\, = 3.6$~GeV instead
of 1.0 and 3.35~GeV.} are given in~\ct{otune-hwar}. The errors of the 
Monte Carlo predictions are comparable to those of the
data.

\section{Results\label{sec:results}}

\subsection{Factorial moments\label{subsec:facmom} }

Fig.~\ref{fig_fpt}  shows cylindrically cut factorial moments $F_q$ of
order $q=2$ to 5 as a function of $\ptc$.\footnote{
 The numerical values of the data on factorial moments and cumulants will 
be made available in the Durham HEP Database, 
{\tt http://durpdg.dur.ac.uk/HEPDATA}~.} 
With decreasing $\ptc$, the  moments decrease towards a minimum at a
common value of $\ptc \approx 1$~GeV but remain larger than unity, the
Poisson value. The observed  deviation 
 of the $\ptc$-dependence 
from the Poissonian behaviour
 for large  $\ptc$ values
agrees qualitatively with the theoretical expectation
discussed in Sect.~\ref{analysis}. However, for smaller $\ptc$ values the
moments rise strongly, in clear disagreement with the perturbative
QCD result for partons.
 Figure~\ref{fig_fpt} suggests that the predicted Poisson limit for soft
 gluons
 is masked by strong hadronisation effects as $\ptc\to Q_0$.
 The drop of the moments and the characteristic dip  for $\ptc \simeq
1$~GeV indicate, however, that perturbative calculations may be
relevant for hadrons down to a scale of approximately 1~GeV.

The Monte Carlo model calculations 
which include both the parton
cascade and hadronisation, largely follow the trend of the data and,
in particular, reproduce the minimum around $\ptc=1$~GeV.
Differences appear for $\ptc \lssim 1$ GeV, with \HW\ describing the
data better than the models using string fragmentation.
 However, 
the dotted curves, which represent 
\PT\
predictions 
 with the inclusion of
 Bose-Einstein correlations, are in
very good
agreement with the measurements.

To study the influence of resonance decays, event samples were
generated wherein resonances were not allowed to decay. The
enhancement for $\ptc\lssim1$~GeV remained almost unaffected  in the
case of the \PT~model. For the \HW\ model, the suppression of
 the decays led to an increase of the
 factorial moments below
 $\ptc \approx 0.5$~GeV, an effect also observed in \epr\
 studies~\ct{ptpz}.
This increase ranges from a
few percent for $q=2$ to
about
20\% for $q=5$.

Fig.~\ref{fig_fpm}  shows spherically cut factorial moments $F_q$, 
$q=2$ to 5, as a function of $\pc$.
 For large $\pc$ values the moments change very little 
 owing to
 the kinematically limited number of particles per event at high 
 momenta.
 However, for
smaller $\pc$   the moments increase rapidly and show no tendency to
level off for  $\pc\to0$~GeV, contrary to theoretical predictions for
partons. The Monte Carlo models describe the data rather well down
to $\pc\simeq 2-3$~GeV. In that region they are, in fact, very
similar to parton-level (and hadron-level) predictions from \ARI\
shown in~\cite{ptp}. For smaller values, and in sharp contrast to
the data, the  Monte Carlo curves flatten off and remain
approximately constant, a feature also observed in~\cite{ptp}. 
  The influence of Bose-Einstein correlations, as implemented in \PT, 
  is sizeable in that region but insufficient to reproduce the data. 
  Indeed, the LEP measurements \cite{bec-lt} have shown that BEC 
  effect is expected 
  to be larger in the longitudinal direction than that in the transverse
  direction of the jet, 
 though this is not implemented in \PT.

\subsection{Factorial cumulants\label{subsec:faccum} }
The large statistics available in this analysis allows 
 the 
study 
 of the
factorial cumulants, $\kptc$ and $\kpc$, defined in (\ref{kvfv}).
These are a direct measure of the genuine correlations among hadrons
and thus
 present the information in a way which can more readily be interpreted. 
  Fig.~\ref{fig_k34} shows the
cylindrically cut
 and spherically cut
cumulants
of order
$q=3,4$.\footnote{Note that $K_2$ is, by definition, equal to $F_2-1$
and therefore not shown here.}

The cumulant $K_3(\ptc)$ has a similar $\ptc$-dependence 
 to 
the
factorial moments: it is positive at large $\ptc$, decreases towards
a minimum, close to zero around $0.6-0.7$~GeV and rises rapidly as
$\ptc$ is further decreased. Interestingly, the minimum occurs at a
smaller $\ptc$ value than that for $K_2(\ptc)$ (or $F_2-1$,
Fig.~\ref{fig_fpt}) and for the higher-order factorial moments.
Four-particle correlations, measured by $K_4(\ptc)$, are compatible
with zero, within errors. 
 From the rapid increase of $K_2(\ptc)$
and $K_3(\ptc)$ as $\ptc\to0$~GeV, we can conclude that the strong rise
of the factorial moments is predominantly due to genuine particle
correlations.

 Figure~\ref{fig_k34}
 shows that the cumulants in spherically cut phase space
 are significantly larger than the $K_q(\ptc)$ cumulants for $\pc$
 smaller than a few GeV. They continue to increase with decreasing
 $\pc$, a behaviour reflected in the corresponding factorial
 moments in Fig.~\ref{fig_fpm}.

The various Monte Carlo models, shown in Fig.~\ref{fig_k34}, agree
 qualitatively
with the measurements but differ in detail
for all $\ptc$ and $\pc<2$~GeV values.
 The largest
deviations occur for the cumulants $\kpc$ below $\pc\simeq 2$~GeV where
the models start to level off, whereas the measured cumulants
continue to increase as $\pc\to0$~GeV. Suppression of resonance
decays
in  \HW\ was found to increase the cumulants
$\kptc$ for
$\ptc$ below $1$~GeV. In contrast, the spherically-cut cumulants
predicted by \HW\ remain essentially unchanged.

The results presented above might potentially be biased due to
correlations induced by Dalitz decays ($\pi^0\to\gamma\,$\ep) and
fake pairs. The former would manifest themselves as a narrow peak in the
invariant mass distribution of unlike-sign particle pairs near
threshold. No such enhancement was found in the data.
 The check against fake pairs showed a peak at very small invariant
 masses which was found not to influence the results.

 We have also repeated the analysis for multiplets composed of like-sign particles.
Although the corresponding moments differ in magnitude, their
dependence on $\ptc$ and $\pc$ (not shown) follows closely that of
all-charge particle moments. We may therefore conclude that the
dip-structure observed in Fig.~\ref{fig_fpt} and Fig.~\ref{fig_k34}
is stable against changes in the charge composition of the
multiplets.

\subsection{Comparison with deep-inelastic scattering\la{zeus} }
The ZEUS measurements reported in~\cite{ptpz} were carried out in
the current region of the Breit frame of reference which is
traditionally considered to be the equivalent of a single
event-hemisphere in \ep\ annihilation~\cite{Br}. 
A sample of
\epr\ interactions was used with an average four-momentum transfer
squared of $\al Q^2 \ar \simeq2070 $~GeV$^2$. This corresponds to an
equivalent \ep\ c.m.s. energy, $\sqrt{s}$, of $44$~GeV.

The factorial moments studied in~\cite{ptpz} show many of the
characteristics also reported here. However, an interesting
difference is observed  in the $\ptc$ dependence of the factorial
moments. The distinctive minimum seen in the OPAL data for
$\ptc\approx 1$~GeV, which could signal the borderline between
perturbative and non-perturbative dynamics, is absent in the ZEUS
measurements. The latter remain constant down to $\ptc=1$~GeV below
which value they increase rapidly. The authors interpret their
measurements as the first indication that perturbative QCD fails on
a qualitative level to describe the hadronic multiplicities and that
hadronisation causes a violation of the LPHD hypothesis for
many-particle inclusive observables.

To try and understand the differences between the \ep\ and
\epr\ results, we have carried out further studies to mimic the
ZEUS experimental conditions. These differ significantly from those
in  \ep\ annihilation: (i) the selected current region of the
Breit frame excludes a large part of the central rapidity region in
the $\gamma^\star$-proton rest-frame, the equivalent of the c.m.s
frame in   \ep,  which is included in our analysis; (ii) the rate
of hard jet production is lower in \epr\ collisions at HERA
energies
than in \ep\ at LEP.

To study the influence of the central rapidity region and  the hard
jet
production  rate, we have repeated the analysis for rapidity
intervals
$y>y_0$ with $y_0\geq 1$ and  for two-jet events, the latter
selected
through a cut on the thrust value.

As an example, Fig.~\ref{fig_fpt15s} shows the moments $\fptc$ for
$y_0=1.5$. Omitting the central region in the OPAL data clearly has
a strong effect on the magnitude and $\ptc$-dependence of the
moments. In particular, the minimum around $\ptc=1$~GeV, seen in the
full sample ($y>0$) is absent when only hadrons with $y>1.5$ are
selected. The ZEUS and OPAL data ($y>1.5$) overlap for $q=2$ in the
full $\ptc$ region; higher order moments are very similar below
$\ptc$ of about 1~GeV but differ in magnitude at higher $\ptc$.

In Figure~\ref{fig_fpt15s} we 
 also 
show results for
two-jet events. These events were selected by requiring the thrust
value of an event to be larger than $0.96$. The sample corresponds
to about $30\%$ of all multi-hadronic events. An alternative two-jet
selection based on the Durham jet-finder~\cite{Durham} led to very
small differences which are included in the systematic
uncertainties.

Selecting two-jet events reduces significantly the values of the
factorial moments (and cumulants, not shown) for all $\ptc$ in
comparison with those in the inclusive sample. The OPAL measurements
are lower than the ZEUS data but,
as expected, show qualitatively
the same
behaviour: constant down to $\ptc\approx 1$~GeV followed by a strong
rise as $\ptc\to0$~GeV, with no evidence for a distinct minimum near
$1$~GeV.

The OPAL results on rapidity-restricted factorial moments suggest
that the often assumed equivalence of a single event-hemisphere in
\ep\ annihilation with the current region in the Breit frame for
deep-inelastic \epr\ interactions has to be treated with caution. In
theoretical predictions for QCD cascades, which apply to the whole
jet and moreover focus on soft gluon emissions, characteristic
signatures of soft particle emission, such as colour coherence, are
seen only if particles with small rapidities are also included.
Likewise, selecting two-jet events may introduce a bias which can
mask the effect under study. Our findings 
 suggest an explanation for
the
differences obtained in the ZEUS analysis~\ct{angul} of 
 angular correlations
compared to the LEP
results~\ct{angulee}.

\section{Summary and conclusions\label{sec:summary}}
Analytical perturbative QCD calculations, in agreement with
parton-level
Monte Carlo calculations, show that gluons produced in a jet become
uncorrelated when the gluon transverse momentum relative to the jet
axis, $p_T$,  is restricted to small values. The approach to a
Poisson regime is a direct consequence of colour coherence, or
angular ordering of gluon emissions in the QCD cascade and is
expected to hold also for soft hadrons if Local Parton-Hadron
Duality (LPHD) is valid.

In this paper, the predicted QCD colour coherence effect has been
tested in \ep\ annihilation at the \z0-resonance  using factorial moments 
and
factorial cumulants of the multiplicity distribution of hadrons with
restricted transverse momenta $p_T<\ptc$ or restricted absolute
momenta $p\equiv |{\bf p}|<\pc$.  The analysis is based on a data
sample of about four million events recorded with the \OP\ detector
at LEP.

For cylindrically cut phase space, the factorial and cumulant
moments are predicted 
 by analytical QCD calculations
to reach limiting values close to unity and
zero, respectively,
for small $\ptc$.
 Likewise, in spherically cut phase space,
the 
 factorial and cumulant
moments should saturate at values well above unity and zero,
respectively,
for $\pc\to0$~GeV.
These expectations are not borne out by
the measurements: for $\ptc\lssim1$~GeV or for values of $\pc$ below
a few GeV, the moments and cumulants rise strongly with decreasing
$\ptc$ or $\pc$. QCD based Monte Carlo models which include
hadronisation, reproduce well the change in correlation pattern but
serious
discrepancies remain in the very low momentum region. However,
Bose-Einstein correlations, as implemented in \PT, significantly
improve the agreement between data and Monte Carlo predictions.

 Interestingly, in the region of
large to intermediate $\ptc$, a minimum around $1$~GeV is observed
in the $\ptc$ dependence of the factorial moments,  with a
corresponding minimum, close to zero,  at $\ptc\simeq0.6$ GeV for
the cumulant $K_3$,
  as expected from
  colour coherence.
 One may
interpret the 
 intermediate
 $\ptc$-range $0.6-1.0$~GeV as a borderline between a
regime of perturbative dynamics where the LPHD hypothesis is
justified, and a regime dominated by strong confinement forces which
leads to violation of LPHD for many-particle inclusive observables.

 In a
similar analysis in deep inelastic \epr\ scattering at HERA, no
evidence was found for a Poisson-like regime in cylindrically-cut
phase space. The results presented here show that the characteristic
decrease towards a minimum around $\ptc$ of 1~GeV disappears if
hadrons produced in the central region of rapidity ($y\lssim1$) are
excluded, or if the analysis is restricted to two-jet events. The
former suggests that, for soft particle production, the often
assumed equivalence of a single event-hemisphere in \ep\
annihilation with the current region in the Breit frame of  a deep
inelastic \epr\ collision may be misleading.

\section*{Acknowledgements}

 We thank Wolfgang Ochs for helpful discussions.
 We particularly wish to thank the SL Division for the efficient operation
of the LEP accelerator at all energies
 and for their close cooperation with
our experimental group.  In addition to the support staff at our own
institutions we are pleased to acknowledge the  \\
Department of Energy, USA, \\
National Science Foundation, USA, \\
Particle Physics and Astronomy Research Council, UK, \\
Natural Sciences and Engineering Research Council, Canada, \\
Israel Science Foundation, administered by the Israel
Academy of Science and Humanities, \\
Benoziyo Center for High Energy Physics,\\
Japanese Ministry of Education, Culture, Sports, Science and
Technology (MEXT) and a grant under the MEXT International
Science Research Program,\\
Japanese Society for the Promotion of Science (JSPS),\\
German Israeli Bi-national Science Foundation (GIF), \\
Bundesministerium f\"ur Bildung und Forschung, Germany, \\
National Research Council of Canada, \\
Hungarian Foundation for Scientific Research, OTKA T-038240, 
and T-042864,\\
The NWO/NATO Fund for Scientific Research, the Netherlands.\\


\nwp


 \bef
 \bec
  \epss=18cm
\vspace*{-1.4cm}
\hspace*{-1.2cm}
  \epsf[5 140 540 700]{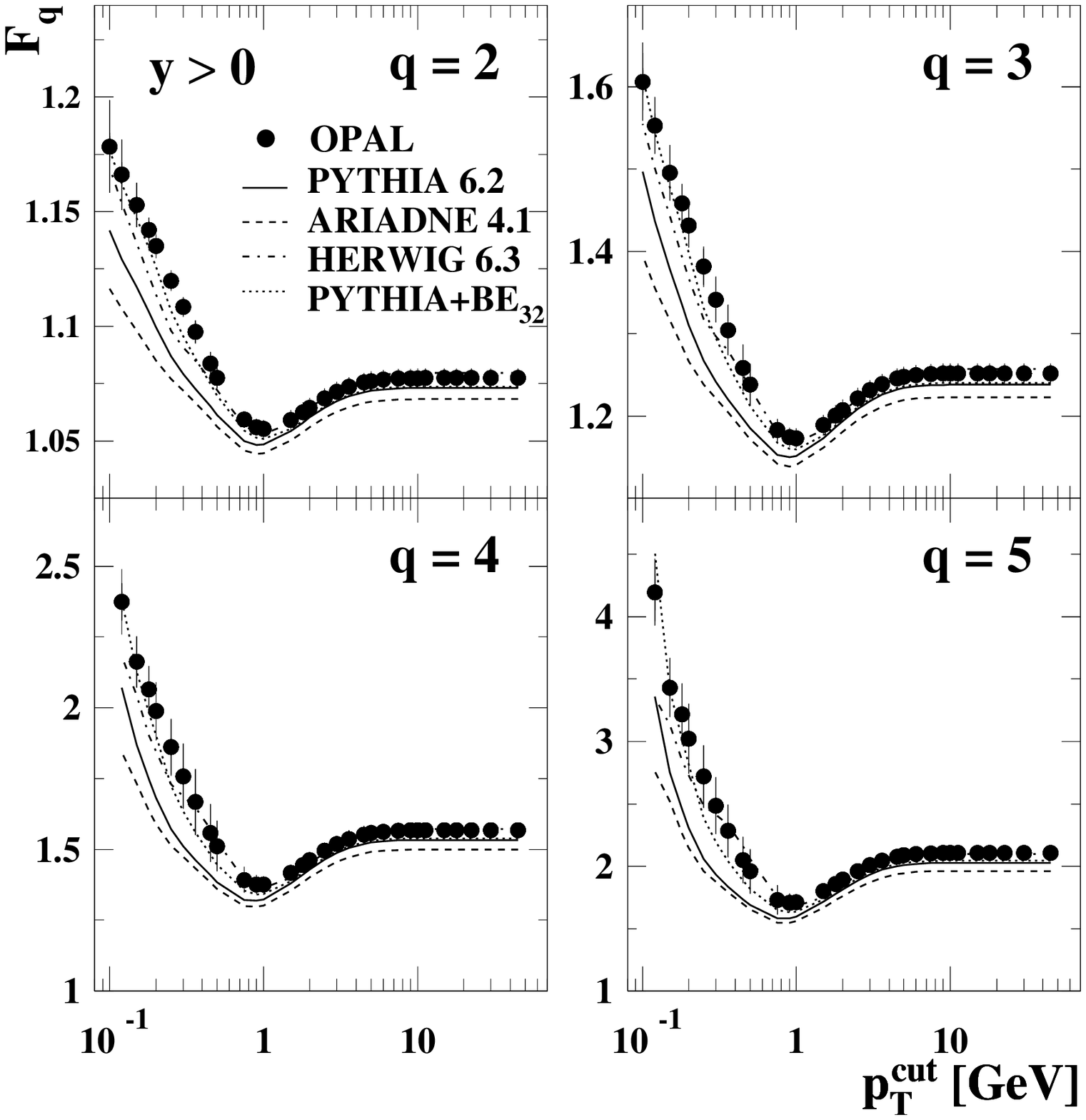}
 \ece
\cptn{\it Factorial moments of charged particles with  transverse
momenta $\pT<\ptc$ as a function of $\ptc$ compared to different
Monte Carlo predictions. \errfigcap 
 \becfigcap 
 \mcerfigcap}
\la{fig_fpt}
\efi

 \bef
 \bec
  \epss=18cm
\vspace*{-1.4cm}
\hspace*{-1.2cm}
  \epsf[5 140 540 700]{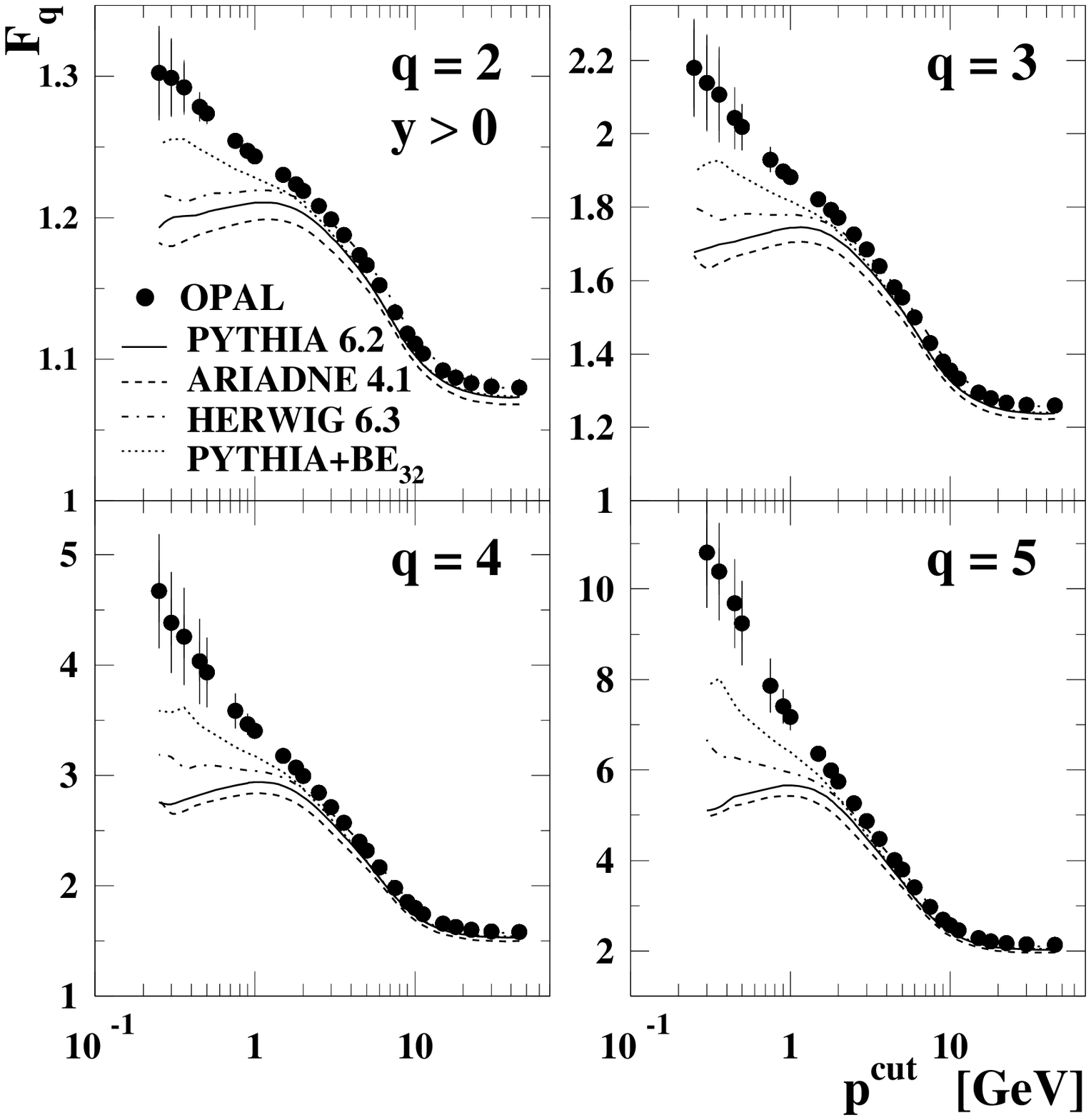}
 \ece
\cptn{\it Factorial moments of charged particles
with absolute
momenta  $p\equiv |{\bf p}|<\pc$ as a function of the
the momentum cut, $\pc$
compared to different Monte Carlo
predictions.
 \errfigcap 
 \becfigcap
 \mcerfigcap
}
\la{fig_fpm}
\efi

 \bef
 \bec
  \epss=18cm
\vspace*{-1.4cm}
\hspace*{-1.2cm}
  \epsf[5 140 540 700]{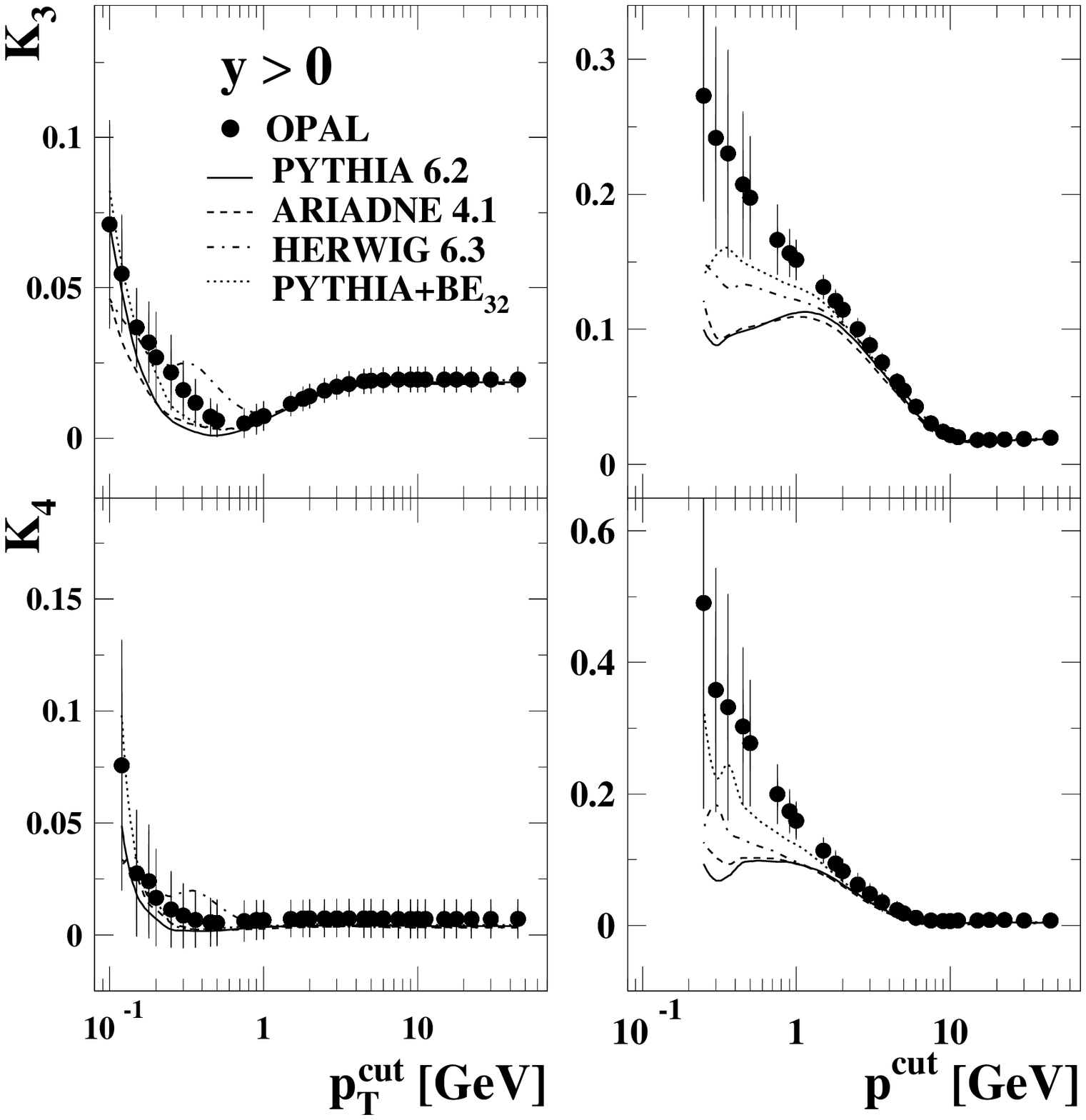}
 \ece
\cptn{\it $K_3$ and $K_4$ cumulants
of charged particles  with transverse
momenta  $\pT<\ptc$
as a function
of
the momentum cut $\ptc$ (left panel) and
with absolute
momenta  $p<\pc$ as a function of the
the momentum cut $\pc$ (right panel)
compared to different Monte Carlo
predictions.
 \errfigcap 
 The errors are correlated bin-to-bin.
 \becfigcap
\mcerfigcap
}
\la{fig_k34}
\efi


 \bef
 \bec
  \epss=18cm
\vspace*{-1.4cm}
\hspace*{-1.2cm}
  \epsf[5 140 540 700]{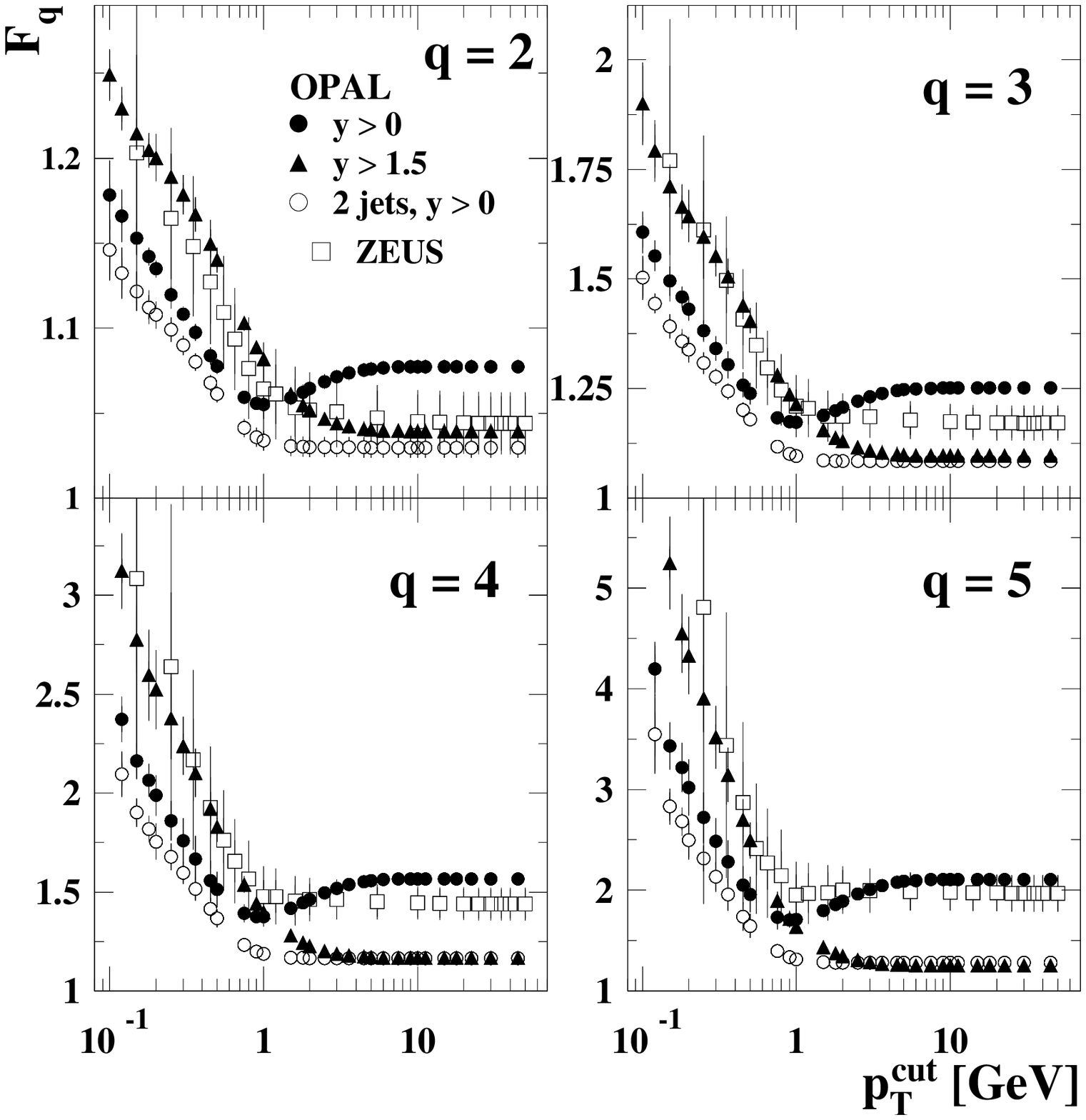}
 \ece
\cptn{\it Factorial moments of charged particles  with absolute
momenta  $\pT<\ptc$ as a function of $\ptc$, compared to those of
2-jet events, in the rapidity window $y>1.5$ and data from
ZEUS~\cite{ptpz}.
\errfigcap
 }
\la{fig_fpt15s} \efi

\end{document}